\documentclass[
showpacs,
floatfix,
aps,
prl,
twocolumn,
superscriptaddress,
amssymb,
]{revtex4-1}

\usepackage{times}
\usepackage{amssymb}
\usepackage{latexsym}
\usepackage[dvips]{graphicx}
\usepackage{amsmath}

\usepackage{graphicx}
\usepackage{dcolumn}
\usepackage{amsfonts}
\usepackage{bm}
\usepackage{epsfig}

\newcommand{\be}{\begin{equation}}
\newcommand{\ee}{\end{equation}}
\newcommand{\bea}{\begin{eqnarray}}
\newcommand{\eea}{\end{eqnarray}}

\def\edc{\epsilon_j}

\def\oc{\omega_{\mbox{\scriptsize {c}}}}

\def\adc{\alpha}

\def\rc{R_{\mbox{\scriptsize {c}}}}

\def\tpi{\tau_{\pi}}

\def\tq{\tau_{\mbox{\scriptsize {q}}}}
\def\ttr{\tau}

\def\tsh{\tau_{\rm sh}}
\def\tsm{\tau_{\rm sm}}
\def\tee{\tau_{\rm ee}}
\def\tin{\tau_{\rm in}}

\def\tst{\tau_\star}

\def\lp{\left (}
\def\rp{\right )}

\newcommand{\req}[1]{Eq.\,(\ref{#1})}

\newcommand{\reqs}[2]{Eqs.\,(\ref{#1}),\,(\ref{#2})}
\newcommand{\rfig}[1]{Fig.\,\ref{#1}}

\newcommand{\rref}[1]{Ref.\,\onlinecite{#1}}

\begin{document}
\title{Nonlinear response of a MgZnO/ZnO heterostructure close to zero bias}
\author{Q.~Shi}
\affiliation{School of Physics and Astronomy, University of Minnesota, Minneapolis, Minnesota 55455, USA}
\author{J.~Falson}
\affiliation{Max-Planck-Institute for Solid State Research, Heisenbergstrasse 1, D-70569 Stuttgart, Germany}
\author{M.~A.~Zudov}
\email[Corresponding author: ]{zudov@physics.umn.edu}
\affiliation{School of Physics and Astronomy, University of Minnesota, Minneapolis, Minnesota 55455, USA}
\author{Y.~Kozuka}
\affiliation{Department of Applied Physics and Quantum-Phase Electronics Center, University of Tokyo, Tokyo 113-8656, Japan}
\author{A.~Tsukazaki}
\affiliation{Institute for Materials Research, Tohoku University, Sendai 980-8577, Japan}
\author{M.~Kawasaki}
\affiliation{Department of Applied Physics and Quantum-Phase Electronics Center, University of Tokyo, Tokyo 113-8656, Japan}
\affiliation{RIKEN Center for Emergent Matter Science, Wako 351-0198, Japan}
\author{J.~Smet}
\affiliation{Max-Planck-Institute for Solid State Research, Heisenbergstrasse 1, D-70569 Stuttgart, Germany}

\begin{abstract}
We report on magnetotransport properties of a MgZnO/ZnO heterostructure subjected to weak direct currents. 
We find that in the regime of overlapping Landau levels, the differential resistivity acquires a quantum correction proportional to both the square of the current and the Dingle factor.
The analysis shows that the correction to the differential resistivity is dominated by a current-induced modification of the electron distribution function and allows us to access both quantum and inelastic scattering rates.
\end{abstract}
\received{26 June 2017}
%\pacs{73.43.Qt, 73.63.Hs, 73.40.-c}
\maketitle

Nonlinear magnetotransport in high Landau levels of two-dimensional electron systems (2DESs) offers a unique approach to obtain information on both electron-impurity and electron-electron scattering.
For example, at high direct currents the differential resistance exhibits Hall field-induced resistance oscillations (HIRO) \citep{yang:2002,bykov:2005c,zhang:2007a,zhang:2008,hatke:2009c,hatke:2010a,hatke:2011a,hatke:2012d,bykov:2012,shi:2014b,shi:2017b} which originate from electron (or hole \cite{shi:2014b}) backscattering off impurities leading to transitions between Landau levels.
Since such transitions are accompanied by a displacement of the electron guiding center by a cyclotron diameter $2\rc$, applied current density $j$ translates to an energy scale $e \rho_H j (2\rc)$, where $\rho_H$ is the Hall resistivity.
HIRO then result from the commensurability between this energy and the inter-Landau level spacing $\hbar\oc$, where $\oc$ is the cyclotron frequency of a charge carrier.
In overlapping Landau levels, the corresponding correction to the differential resistance $r$ is given by \citep{vavilov:2007}
\be
\frac {\delta r}{R_0} \approx \frac{16}{\pi}\frac{\tau}{\tau_{\pi}}\lambda^2 \cos 2\pi \edc \,,~~ \pi\edc \gg 1\,,
\label{eq.hiro}
\ee
where $\edc = e \rho_H j (2\rc)/\hbar\oc$, $R_0$ is the low-temperature, linear-response resistance at zero magnetic field ($B=0$), $\ttr$ is the disorder-limited transport scattering time, $\tpi$ is the backscattering time, $\lambda = \exp(-\pi/\oc\tq)$ is the Dingle factor, and $\tq$ is the quantum lifetime.

The disorder in a 2DES can often be conveniently separated into a short-range (e.g., background impurities) and a long-range (e.g., remote ionized donors) component, characterized by ``sharp'' and ``smooth'' scattering rates ($\tsh^{-1}$ and $\tsm^{-1}$), respectively. 
When $\ttr \gg \tq$, as in a conventional high-mobility modulation-doped 2DES, $\tsh \approx \tpi$ and $\tsm \approx \tq$.
Therefore, the analysis of the HIRO amplitude using \req{eq.hiro} can yield information on both sharp and smooth disorder components in a 2DES under study.

In the regime of weak electric fields, the differential resistance acquires a negative quantum correction which scales with $j^2$, as has been observed in GaAs heterostructures \cite{zhang:2007a,bykov:2007,zhang:2007b,zhang:2008,zhang:2009,vitkalov:2009,hatke:2010a,bykov:2010c,wiedmann:2011c,gusev:2011,hatke:2012d}.
In contrast to \req{eq.hiro}, this current-induced correction originates \emph{both} from the low $\edc$ counterpart of \req{eq.hiro} \citep{vavilov:2007} (displacement mechanism) and from the oscillatory modification of the energy distribution function (inelastic mechanism) \citep{dmitriev:2005}.
More specifically, in overlapping Landau levels, $\delta r$ can be written as \citep{vavilov:2007}
\be
\frac{\delta r}{R_0} \approx - \alpha\edc^2\,, ~~ \pi\edc \ll \min\{1,(2\ttr/\tin)^{1/2}\}\,,
\label{eq.adc.1}
\ee
where
\be
\alpha = \alpha_0\lambda^{2}\,,~ \adc_0 = 12 \pi^2\lp \frac {3\ttr}{16\tst} + \frac {\tin}{\ttr} \rp\,.
\label{eq.adc.2}
\ee
Here, $\tst^{-1}$ entering the first (displacement) term can be expressed as $\tst^{-1}=3\tau_0^{-1}-4\tau_1^{-1}+\tau_2^{-1}$, where $\tau_n^{-1}$ represents $n$-th angular harmonic of the rate of scattering on angle $\theta$, $\tau_\theta^{-1} = \sum\tau_n^{-1}e^{i n\theta}$ \citep{note:51}.
This displacement term can never exceed $9/16$ (sharp-disorder limit).
In contrast, the second (inelastic) term in \req{eq.adc.2}, given by $\tin/\ttr \sim \hbar E_F/\ttr(k_B T)^2$ ($E_F$ is the Fermi energy, $\tin$ is the inelastic relaxation time), can be significantly larger than unity, especially in high density and low mobility 2DESs \cite{bykov:2007,zhang:2007b,zhang:2009,vitkalov:2009,bykov:2010c,wiedmann:2011c,gusev:2011,hatke:2012d}.
In such systems, nonlinear transport at small $j$ offers a convenient way to obtain $\tin$ and thus access the strength of electron-electron interactions in the 2DES under study.

In this paper the capability of nonlinear transport to reveal information about scattering sources is exploited on a Mg$_x$Zn$_{1-x}$O/ZnO heterostructure \citep{kozuka:2014,falson:2015,falson:2015b,kozlov:2015,falson:2016,karcher:2016,shi:2017b}.
We find that the correction to the differential resistivity can be well described by \req{eq.adc.1}.
The analysis of the curvature $\alpha$ reveals that the observed nonlinear response is governed by the current-induced modification of the electron distribution function, consistent with the theoretical estimates.
From the Dingle analysis we obtain $\tq \approx 2$ ps, in good agreement with the values found from recent measurements of Shubnikov-de Haas \citep{falson:2015b}, microwave-induced \citep{karcher:2016}, and Hall field-induced \cite{shi:2017b} resistance oscillations, confirming the applicability of \reqs{eq.adc.1}{eq.adc.2}.
More importantly, our experiments allow us to estimate the inelastic relaxation time $\tin \approx 40$ ps, which could not be accessed in previous studies \citep{falson:2015b,karcher:2016,shi:2017b}.
While a similar approach has been previously employed to study inelastic relaxation in GaAs/AlGaAs quantum wells, a much lower mobility and higher density of our MgZnO/ZnO heterostructure suggest its potential applicability to a diverse variety of 2DESs.

Our sample was fabricated from a Mg$_x$Zn$_{1-x}$O/ZnO heterostructure grown using liquid ozone-based molecular beam epitaxy \cite{falson:2011,falson:2016}.
A Hall bar of width of $\approx 0.09$ mm and distance between voltage probes of $\approx 0.8$ mm was defined by scratching the wafer with a tungsten needle \citep{shi:2017b}.
Electrical contacts were made by soldered indium.
At $T \approx 1.35$ K, our 2DES has density $n_e \approx 2.0 \times 10^{12}$ cm$^{-2}$ and mobility $\mu \approx 2.3 \times 10^4$ cm$^2$/Vs.
The differential resistance $r$ was recorded using a standard four-terminal lock-in technique at a constant coolant temperature $T \approx 1.35$ K either while sweeping magnetic field $B$ at constant direct current $I$ or while sweeping $I$ at a constant $B$.

%%%%%%%%%%%%%%%%%%%%%%%%%%%%%%%%%
\begin{figure}
\includegraphics{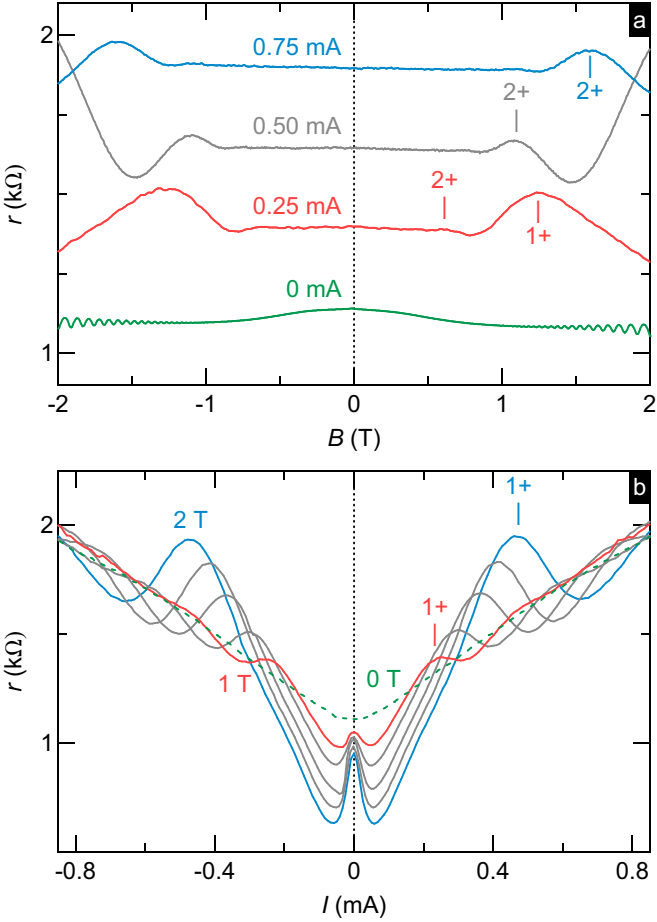}
\vspace{-0.1 in}
\caption{(Color online)
(a) $r(B)$ at different $I$ from 0 (bottom curve) to 0.75 mA (top curve), in steps of 0.25 mA.
HIRO maxima are marked by $1+$ and $2+$.
The curves are not shifted.
(b) $r(I)$ at different $B$ from 1 to 2 T, in steps of 0.25 T. 
The dashed line represents $r(I)$ at $B=0$.
}
\vspace{-0.1 in}
\label{fig1}
\end{figure}
%%%%%%%%%%%%%%%%%%%%%%%%%%%%%%%%%%
To facilitate the discussion of our results in the regime of weak currents, we first briefly summarize the main findings of related HIRO experiments.
In \rfig{fig1}(a) we present $r$ as a function of $B$ recorded at different $I$ from 0 (bottom curve) to 0.75 mA (top curve), in steps of 0.25 mA.
The trace at $I = 0$ is rather featureless, showing only weak Shubnikov-de Haas oscillations at $B \gtrsim 1$ T.
The data at $I = 0.25$ mA and higher currents, however, reveal HIRO (cf. $1+,2+$) which spread over a wider $B$ range with increasing $I$.
The positions of the HIRO maxima are well described by $B_N^+ \approx ({m^\star}\sqrt{{8\pi}/{n_e}}/{e^2})(j/N)\propto j/N$ \citep{shi:2017b}, where $N = 1,2,3,...\,$.
However, in contrast to what one might expect, the increase in $I$ does not lead to observation of more oscillations.
This is in line with the recent study \citep{shi:2017b} which found that $\tq$ decreases rapidly with $I$.
Also, like \rref{shi:2017b}, we observe that the zero-field differential resistance $r_0$ also increases with $I$, suggesting a decrease of $\ttr$.
Both observations are consistent with a scenario that Joule heating leads to elevated electron temperature which, in turn, causes enhanced electron-electron and electron-phonon scattering \citep{shi:2017b}.
Owing to this unintentional heating, our HIRO experiments never revealed more than three oscillations (regardless of $I$) which precluded extracting $\tq$ from a conventional Dingle analysis.
Instead, we had to resort \citep{shi:2017b} to fitting experimental curves with the HIRO expression whose applicability, in contrast to \req{eq.hiro}, is not limited to $\pi\edc \gg 1$ \citep{vavilov:2007}:
\be
\frac {\delta r}{R_0} = -\frac{2\tau}{\tsh} \lambda^2 \left ( \zeta  \left [ J_0^2(\zeta) \right ]'' \right )'\,.
\label{eq.bessel}
\ee
Here, $J_0$ is the Bessel function and the prime denotes a derivative with respect $\zeta = \pi \edc$.
Not surprisingly, the obtained $\tq$ value showed significant dependence on $I$, reflecting a sizable electron-electron scattering contribution \citep{hatke:2009a,hatke:2009c}.
As a result, the impurity-limited $\tq$ could only be estimated by extrapolating the $I$-dependence of the quantum scattering rate to zero current \citep{shi:2017b}.

An alternative way to study HIRO is to sweep $I$ while keeping $B$ fixed \citep{zhang:2007a}.
In \rfig{fig1}(b) we present $r(I)$ at different $B$ from 1 to 2 T in steps of 0.25 T (solid lines) and at $B=0$ (dashed line).
At $B = 1$ T and higher, oscillations in $r$ with $I$ are superimposed on a smooth, monotonically increasing background which closely mimics $r(I)$ at $B = 0$.
Concurrently, the oscillations move to higher $I$ with increasing $B$ while growing in amplitude.
This increase originates from enhanced modulation of the density of states at higher $B$.

At the focus of the present work is a maximum centered at $I = 0$ which becomes more pronounced with increasing $B$, see \rfig{fig1}(b). 
As we show below, this maximum appears due to Landau quantization, primarily, as a result of current-induced modification of the electron distribution function \citep{dmitriev:2005,vavilov:2007}.
We further illustrate how this feature can be used to obtain both quantum and inelastic lifetimes.

%%%%%%%%%%%%%%%%%%%%%%%%%%%%%%%%%%%%%%%%%%%%
\begin{figure}
\includegraphics{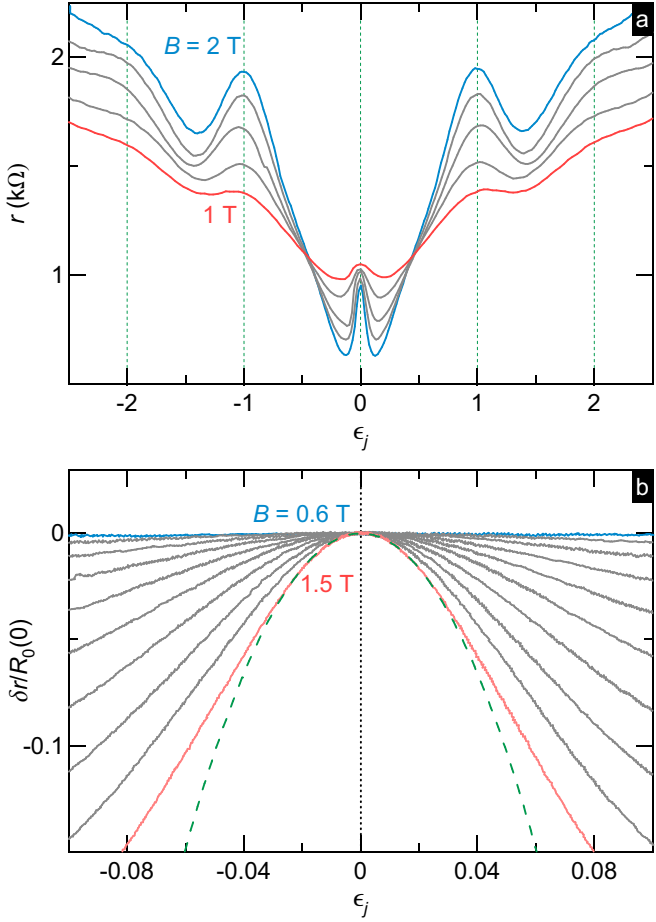}
\vspace{-0.1 in}
\caption{(Color online) 
(a) $r(\edc)$ at different $B$ from 1 to 2 T, in steps of 0.25 T. 
(b) $\delta r/R_0(0)$ vs. $\edc$ at different $B$ from 0.6 to 1.5 T, in steps of 0.1 T.
The dashed line is the fit to the data at $B = 1.5$ T with \req{eq.adc.1}. 
}
\vspace{-0.1 in}
\label{fig2}
\end{figure}
%%%%%%%%%%%%%%%%%%%%%%%%%%%%%%%%%

In \rfig{fig2}(a) we replot $r$ shown in \rfig{fig1}(b) as a function of $\edc = 2e \rho_H\rc j/\hbar\oc$, where $\oc = eB/m^\star$, $m^\star = 0.3 m_0$ \citep{shi:2017b}, and $m_0$ is the mass of a free electron.
Oscillations at all $B$ are lined up and display maxima at $\edc \approx 1, 2$ and a minimum at $\edc \approx 1.5$, indicating that $\edc$ was obtained consistently.
The reduction of the oscillation amplitude at smaller $B$ is as result of increased Landau level overlap manifested in the decay of the Dingle factor.
We also notice that the oscillation amplitude decreases with $\edc \propto j$ which can be attributed to Joule heating discussed above.

At small $\edc^2 \ll 1$, the differential resistance is suppressed and this suppression becomes more pronounced with increasing $B$.
To examine this regime in detail, we present in \rfig{fig2}(b) $\delta r/R_0(0) \equiv [r-r(0)]/R_0(0)$ as a function of $\edc$ at different $B$ from 0.6 (top) to 1.5 T (bottom), in steps of 0.1 T.
While at $B = 0.6$ T the trace is virtually flat, the curvature rapidly increases with $B$.
An example of a fit to the data at $B = 1.5$ T with \req{eq.adc.1} (dashed line) demonstrates excellent overlap with the experimental data at $|\edc| \lesssim 0.02$.
As we show below, the applicability condition of \req{eq.adc.1} is well satisfied since $(2\ttr/\tin)^{1/2} \approx 0.5$, which far exceeds our fitting range of $|\pi\edc| \lesssim 0.06$

%%%%%%%%%%%%%%%%%%%%%%%%%%%%%%%%%%%%%%%%%%%%
\begin{figure}[t]
\includegraphics{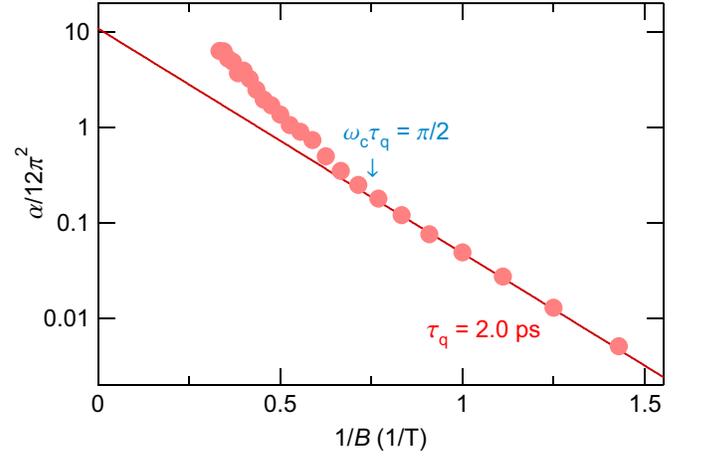}
\vspace{-0.1 in}
\caption{(Color online) 
Reduced curvature $\alpha/12\pi^2$, obtained from the fits in \rfig{fig2}(b) as a function of $1/B$ on a log-linear scale.
The fit to the lower $B$ part of the data with \req{eq.adc.2} (solid line) generates $\tq \approx 2.0$ ps and $3\ttr/16\tst + \tin/\ttr \approx 11$. 
}
\vspace{-0.1 in}
\label{fig3}
\end{figure}
%%%%%%%%%%%%%%%%%%%%%%%%%%%%%%%%%

We next fit our data at all other $B$ and obtain the curvature $\alpha$ which is the only fitting parameter.
Being guided by \req{eq.adc.2}, we construct a Dingle plot, presented in \rfig{fig3}, showing the reduced curvature $\alpha/12\pi^2$ as a function of $1/B$ on a log-linear scale. 
At $1/B \gtrsim 0.7$ T, the curvature exhibits exponential dependence from which we obtain $\tq \approx 2.0$ ps using \req{eq.adc.2}. 
The deviation of the data from the exponential dependence at lower $1/B$ has been previously observed in experiments on GaAs quantum wells in the regime of separated Landau levels \citep{hatke:2012d}.
In our MgZnO/ZnO sample, the Landau levels start to separate at $B \approx 1.3$ T, as estimated from $\oc\tq  = \pi/2$ \citep{ando:1974b,laikhtman:1994}.

We now examine the intercept of the Dingle plot in \rfig{fig3} and extract the inelastic scattering time. 
We first recall that the first term in \req{eq.adc.2}, representing the displacement contribution, has its maximal value of $3\tau/16\tst \approx 9/16$ (sharp disorder limit), a condition which was established in a recent HIRO study \citep{shi:2017b}. 
Since the intercept of the fit in \rfig{fig3} yields $\alpha_0/12\pi^2 \approx 11 \gg 9/16$, we conclude that the inelastic mechanism dominates the nonlinear response at small direct currents in our MgZnO/ZnO heterostructure.
From this value, we then find $\tin \approx 40$ ps.

It is interesting to compare the obtained value of $\tin$ to the one expected from theoretical considerations \citep{giuliani:1982,dmitriev:2005}.
In the regime of our experiment \citep{note:8}, theory predicts
\be
\tin \approx 0.82 \tee\,,~~ \frac \hbar \tee \approx \frac{\pi k_B^2 T^2}{4 E_F}\ln\left( \frac{2 v_F}{a_B \oc \sqrt{\oc\ttr}}\right)\,,
\label{eq.tee}
\ee
where $\tee^{-1}$ is the electron-electron scattering rate for a test particle at the Fermi energy, $v_F$ is the Fermi velocity, and $a_B \approx 1.5$ nm is the Bohr radius.
At $B = 0.7$ T, the logarithmic factor is about 5.9 and \req{eq.tee} yields $\tin \approx 140$ ps, a few times larger than the value obtained from our Dingle analysis.

Since we have limited our analysis to currents considerably smaller than those needed to induce noticeable change in $r$ at $B = 0$, we believe that any residual Joule heating is unlikely to be responsible for this discrepancy.
On the other hand, \reqs{eq.adc.1}{eq.adc.2} were derived assuming $\oc\ttr \gg 1$ and $\oc\tq \lesssim 1$. 
Since in high-mobility GaAs samples $\ttr \gg \tq$, both of these conditions can be simultaneously met.
The situation is markedly different in our MgZnO/ZnO sample where $\tq \lesssim \ttr \approx 3.8$ ps and the above constraints are only marginally satisfied.
Indeed, at $B = 0.7$ T we estimate $\oc\ttr \approx 1.5$ and $\oc\tq \approx 0.8$.

In summary, we have studied nonlinear magnetotransport in a Mg$_x$Zn$_{1-x}$O/ZnO heterostructure in the regime of weak direct currents.
We have found that the  differential resistivity acquires a correction $\delta r$ which is quadratic in direct current and decays exponentially with $1/B$.
The analysis of the $B$-dependence of the curvature suggests that the nonlinear response is governed by a dc field-induced modification of the electron distribution function.
The Dingle analysis in the regime of overlapping Landau levels reveals the quantum lifetime $\tq \approx 2.0$ ps, consistent with existing magnetotransport studies in similar Mg$_x$Zn$_{1-x}$O/ZnO heterostructures \citep{falson:2015b,karcher:2016,shi:2017b}.
However, the regime of small direct currents explored in the present work also allowed us to obtain the inelastic relaxation time of $\tin \approx 40$ ps, which was not previously measured in Mg$_x$Zn$_{1-x}$O/ZnO heterostructures.
Our study demonstrates that the nonlinear magnetotransport is a powerful technique to access scattering times in 2D systems not necessarily having high mobility.
As such, the technique could be useful to investigate electron-impurity and electron-electron scattering in a broad variety of materials.

\begin{acknowledgments}
We thank I. A. Dmitriev for discussions and P. Herlinger for assistance with the microscope.
This work was supported by the US Department of Energy, Office of Basic Energy Sciences, under Grant No. DE-SC002567 (University of Minnesota) and by Grant-in-Aids for Scientific Research (S) No. 24226002 from MEXT, Japan (University of Tokyo). 
\end{acknowledgments}


\begin{thebibliography}{33}
\expandafter\ifx\csname natexlab\endcsname\relax\def\natexlab#1{#1}\fi
\expandafter\ifx\csname bibnamefont\endcsname\relax
  \def\bibnamefont#1{#1}\fi
\expandafter\ifx\csname bibfnamefont\endcsname\relax
  \def\bibfnamefont#1{#1}\fi
\expandafter\ifx\csname citenamefont\endcsname\relax
  \def\citenamefont#1{#1}\fi
\expandafter\ifx\csname url\endcsname\relax
  \def\url#1{\texttt{#1}}\fi
\expandafter\ifx\csname urlprefix\endcsname\relax\def\urlprefix{URL }\fi
\providecommand{\bibinfo}[2]{#2}
\providecommand{\eprint}[2][]{\url{#2}}

\bibitem[{\citenamefont{Yang et~al.}(2002)\citenamefont{Yang, Zhang, Du,
  Simmons, and Reno}}]{yang:2002}
\bibinfo{author}{\bibfnamefont{C.~L.} \bibnamefont{Yang}},
  \bibinfo{author}{\bibfnamefont{J.}~\bibnamefont{Zhang}},
  \bibinfo{author}{\bibfnamefont{R.~R.} \bibnamefont{Du}},
  \bibinfo{author}{\bibfnamefont{J.~A.} \bibnamefont{Simmons}},
  \bibnamefont{and} \bibinfo{author}{\bibfnamefont{J.~L.} \bibnamefont{Reno}},
  \bibinfo{journal}{Phys. Rev. Lett.} \textbf{\bibinfo{volume}{89}},
  \bibinfo{pages}{076801} (\bibinfo{year}{2002}).

\bibitem[{\citenamefont{Bykov et~al.}(2005)\citenamefont{Bykov, Zhang,
  Vitkalov, Kalagin, and Bakarov}}]{bykov:2005c}
\bibinfo{author}{\bibfnamefont{A.~A.} \bibnamefont{Bykov}},
  \bibinfo{author}{\bibfnamefont{J.}~\bibnamefont{Zhang}},
  \bibinfo{author}{\bibfnamefont{S.}~\bibnamefont{Vitkalov}},
  \bibinfo{author}{\bibfnamefont{A.~K.} \bibnamefont{Kalagin}},
  \bibnamefont{and} \bibinfo{author}{\bibfnamefont{A.~K.}
  \bibnamefont{Bakarov}}, \bibinfo{journal}{Phys. Rev. B}
  \textbf{\bibinfo{volume}{72}}, \bibinfo{pages}{245307}
  (\bibinfo{year}{2005}).

\bibitem[{\citenamefont{Zhang et~al.}(2007{\natexlab{a}})\citenamefont{Zhang,
  Chiang, Zudov, Pfeiffer, and West}}]{zhang:2007a}
\bibinfo{author}{\bibfnamefont{W.}~\bibnamefont{Zhang}},
  \bibinfo{author}{\bibfnamefont{H.-S.} \bibnamefont{Chiang}},
  \bibinfo{author}{\bibfnamefont{M.~A.} \bibnamefont{Zudov}},
  \bibinfo{author}{\bibfnamefont{L.~N.} \bibnamefont{Pfeiffer}},
  \bibnamefont{and} \bibinfo{author}{\bibfnamefont{K.~W.} \bibnamefont{West}},
  \bibinfo{journal}{Phys. Rev. B} \textbf{\bibinfo{volume}{75}},
  \bibinfo{pages}{041304(R)} (\bibinfo{year}{2007}{\natexlab{a}}).

\bibitem[{\citenamefont{Zhang et~al.}(2008)\citenamefont{Zhang, Zudov,
  Pfeiffer, and West}}]{zhang:2008}
\bibinfo{author}{\bibfnamefont{W.}~\bibnamefont{Zhang}},
  \bibinfo{author}{\bibfnamefont{M.~A.} \bibnamefont{Zudov}},
  \bibinfo{author}{\bibfnamefont{L.~N.} \bibnamefont{Pfeiffer}},
  \bibnamefont{and} \bibinfo{author}{\bibfnamefont{K.~W.} \bibnamefont{West}},
  \bibinfo{journal}{Phys. Rev. Lett.} \textbf{\bibinfo{volume}{100}},
  \bibinfo{pages}{036805} (\bibinfo{year}{2008}).

\bibitem[{\citenamefont{Hatke et~al.}(2009{\natexlab{a}})\citenamefont{Hatke,
  Zudov, Pfeiffer, and West}}]{hatke:2009c}
\bibinfo{author}{\bibfnamefont{A.~T.} \bibnamefont{Hatke}},
  \bibinfo{author}{\bibfnamefont{M.~A.} \bibnamefont{Zudov}},
  \bibinfo{author}{\bibfnamefont{L.~N.} \bibnamefont{Pfeiffer}},
  \bibnamefont{and} \bibinfo{author}{\bibfnamefont{K.~W.} \bibnamefont{West}},
  \bibinfo{journal}{Phys. Rev. B} \textbf{\bibinfo{volume}{79}},
  \bibinfo{pages}{161308(R)} (\bibinfo{year}{2009}{\natexlab{a}}).

\bibitem[{\citenamefont{Hatke et~al.}(2010)\citenamefont{Hatke, Chiang, Zudov,
  Pfeiffer, and West}}]{hatke:2010a}
\bibinfo{author}{\bibfnamefont{A.~T.} \bibnamefont{Hatke}},
  \bibinfo{author}{\bibfnamefont{H.-S.} \bibnamefont{Chiang}},
  \bibinfo{author}{\bibfnamefont{M.~A.} \bibnamefont{Zudov}},
  \bibinfo{author}{\bibfnamefont{L.~N.} \bibnamefont{Pfeiffer}},
  \bibnamefont{and} \bibinfo{author}{\bibfnamefont{K.~W.} \bibnamefont{West}},
  \bibinfo{journal}{Phys. Rev. B} \textbf{\bibinfo{volume}{82}},
  \bibinfo{pages}{041304(R)} (\bibinfo{year}{2010}).

\bibitem[{\citenamefont{Hatke et~al.}(2011)\citenamefont{Hatke, Zudov,
  Pfeiffer, and West}}]{hatke:2011a}
\bibinfo{author}{\bibfnamefont{A.~T.} \bibnamefont{Hatke}},
  \bibinfo{author}{\bibfnamefont{M.~A.} \bibnamefont{Zudov}},
  \bibinfo{author}{\bibfnamefont{L.~N.} \bibnamefont{Pfeiffer}},
  \bibnamefont{and} \bibinfo{author}{\bibfnamefont{K.~W.} \bibnamefont{West}},
  \bibinfo{journal}{Phys. Rev. B} \textbf{\bibinfo{volume}{83}},
  \bibinfo{pages}{081301(R)} (\bibinfo{year}{2011}).

\bibitem[{\citenamefont{Hatke et~al.}(2012)\citenamefont{Hatke, Zudov,
  Pfeiffer, and West}}]{hatke:2012d}
\bibinfo{author}{\bibfnamefont{A.~T.} \bibnamefont{Hatke}},
  \bibinfo{author}{\bibfnamefont{M.~A.} \bibnamefont{Zudov}},
  \bibinfo{author}{\bibfnamefont{L.~N.} \bibnamefont{Pfeiffer}},
  \bibnamefont{and} \bibinfo{author}{\bibfnamefont{K.~W.} \bibnamefont{West}},
  \bibinfo{journal}{Phys. Rev. B} \textbf{\bibinfo{volume}{86}},
  \bibinfo{pages}{081307(R)} (\bibinfo{year}{2012}).

\bibitem[{\citenamefont{Bykov et~al.}(2012)\citenamefont{Bykov, Dmitriev,
  Marchishin, Byrnes, and Vitkalov}}]{bykov:2012}
\bibinfo{author}{\bibfnamefont{A.~A.} \bibnamefont{Bykov}},
  \bibinfo{author}{\bibfnamefont{D.~V.} \bibnamefont{Dmitriev}},
  \bibinfo{author}{\bibfnamefont{I.~V.} \bibnamefont{Marchishin}},
  \bibinfo{author}{\bibfnamefont{S.}~\bibnamefont{Byrnes}}, \bibnamefont{and}
  \bibinfo{author}{\bibfnamefont{S.~A.} \bibnamefont{Vitkalov}},
  \bibinfo{journal}{Appl. Phys. Lett.} \textbf{\bibinfo{volume}{100}},
  \bibinfo{eid}{251602} (\bibinfo{year}{2012}).

\bibitem[{\citenamefont{Shi et~al.}(2014)\citenamefont{Shi, Ebner, and
  Zudov}}]{shi:2014b}
\bibinfo{author}{\bibfnamefont{Q.}~\bibnamefont{Shi}},
  \bibinfo{author}{\bibfnamefont{Q.~A.} \bibnamefont{Ebner}}, \bibnamefont{and}
  \bibinfo{author}{\bibfnamefont{M.~A.} \bibnamefont{Zudov}},
  \bibinfo{journal}{Phys. Rev. B} \textbf{\bibinfo{volume}{90}},
  \bibinfo{pages}{161301(R)} (\bibinfo{year}{2014}).

\bibitem[{\citenamefont{Shi et~al.}(2017)\citenamefont{Shi, Zudov, Falson,
  Kozuka, Tsukazaki, Kawasaki, von Klitzing, and Smet}}]{shi:2017b}
\bibinfo{author}{\bibfnamefont{Q.}~\bibnamefont{Shi}},
  \bibinfo{author}{\bibfnamefont{M.~A.} \bibnamefont{Zudov}},
  \bibinfo{author}{\bibfnamefont{J.}~\bibnamefont{Falson}},
  \bibinfo{author}{\bibfnamefont{Y.}~\bibnamefont{Kozuka}},
  \bibinfo{author}{\bibfnamefont{A.}~\bibnamefont{Tsukazaki}},
  \bibinfo{author}{\bibfnamefont{M.}~\bibnamefont{Kawasaki}},
  \bibinfo{author}{\bibfnamefont{K.}~\bibnamefont{von Klitzing}},
  \bibnamefont{and} \bibinfo{author}{\bibfnamefont{J.}~\bibnamefont{Smet}},
  \bibinfo{journal}{Phys. Rev. B} \textbf{\bibinfo{volume}{95}},
  \bibinfo{pages}{041411(R)} (\bibinfo{year}{2017}).

\bibitem[{\citenamefont{Vavilov et~al.}(2007)\citenamefont{Vavilov, Aleiner,
  and Glazman}}]{vavilov:2007}
\bibinfo{author}{\bibfnamefont{M.~G.} \bibnamefont{Vavilov}},
  \bibinfo{author}{\bibfnamefont{I.~L.} \bibnamefont{Aleiner}},
  \bibnamefont{and} \bibinfo{author}{\bibfnamefont{L.~I.}
  \bibnamefont{Glazman}}, \bibinfo{journal}{Phys. Rev. B}
  \textbf{\bibinfo{volume}{76}}, \bibinfo{pages}{115331}
  (\bibinfo{year}{2007}).

\bibitem[{\citenamefont{Bykov et~al.}(2007)\citenamefont{Bykov, Zhang,
  Vitkalov, Kalagin, and Bakarov}}]{bykov:2007}
\bibinfo{author}{\bibfnamefont{A.~A.} \bibnamefont{Bykov}},
  \bibinfo{author}{\bibfnamefont{J.-Q.} \bibnamefont{Zhang}},
  \bibinfo{author}{\bibfnamefont{S.}~\bibnamefont{Vitkalov}},
  \bibinfo{author}{\bibfnamefont{A.~K.} \bibnamefont{Kalagin}},
  \bibnamefont{and} \bibinfo{author}{\bibfnamefont{A.~K.}
  \bibnamefont{Bakarov}}, \bibinfo{journal}{Phys. Rev. Lett.}
  \textbf{\bibinfo{volume}{99}}, \bibinfo{pages}{116801}
  (\bibinfo{year}{2007}).

\bibitem[{\citenamefont{Zhang et~al.}(2007{\natexlab{b}})\citenamefont{Zhang,
  Vitkalov, Bykov, Kalagin, and Bakarov}}]{zhang:2007b}
\bibinfo{author}{\bibfnamefont{J.~Q.} \bibnamefont{Zhang}},
  \bibinfo{author}{\bibfnamefont{S.}~\bibnamefont{Vitkalov}},
  \bibinfo{author}{\bibfnamefont{A.~A.} \bibnamefont{Bykov}},
  \bibinfo{author}{\bibfnamefont{A.~K.} \bibnamefont{Kalagin}},
  \bibnamefont{and} \bibinfo{author}{\bibfnamefont{A.~K.}
  \bibnamefont{Bakarov}}, \bibinfo{journal}{Phys. Rev. B}
  \textbf{\bibinfo{volume}{75}}, \bibinfo{pages}{081305(R)}
  (\bibinfo{year}{2007}{\natexlab{b}}).

\bibitem[{\citenamefont{Zhang et~al.}(2009)\citenamefont{Zhang, Vitkalov, and
  Bykov}}]{zhang:2009}
\bibinfo{author}{\bibfnamefont{J.~Q.} \bibnamefont{Zhang}},
  \bibinfo{author}{\bibfnamefont{S.}~\bibnamefont{Vitkalov}}, \bibnamefont{and}
  \bibinfo{author}{\bibfnamefont{A.~A.} \bibnamefont{Bykov}},
  \bibinfo{journal}{Phys. Rev. B} \textbf{\bibinfo{volume}{80}},
  \bibinfo{pages}{045310} (\bibinfo{year}{2009}).

\bibitem[{\citenamefont{Vitkalov}(2009)}]{vitkalov:2009}
\bibinfo{author}{\bibfnamefont{S.}~\bibnamefont{Vitkalov}},
  \bibinfo{journal}{Int. J. Mod. Phys. B} \textbf{\bibinfo{volume}{23}},
  \bibinfo{pages}{4727} (\bibinfo{year}{2009}).

\bibitem[{\citenamefont{Bykov et~al.}(2010)\citenamefont{Bykov, Mozulev, and
  Vitkalov}}]{bykov:2010c}
\bibinfo{author}{\bibfnamefont{A.~A.} \bibnamefont{Bykov}},
  \bibinfo{author}{\bibfnamefont{E.~G.} \bibnamefont{Mozulev}},
  \bibnamefont{and} \bibinfo{author}{\bibfnamefont{S.~A.}
  \bibnamefont{Vitkalov}}, \bibinfo{journal}{JETP Lett.}
  \textbf{\bibinfo{volume}{92}}, \bibinfo{pages}{475} (\bibinfo{year}{2010}).

\bibitem[{\citenamefont{Wiedmann et~al.}(2011)\citenamefont{Wiedmann, Gusev,
  Raichev, Bakarov, and Portal}}]{wiedmann:2011c}
\bibinfo{author}{\bibfnamefont{S.}~\bibnamefont{Wiedmann}},
  \bibinfo{author}{\bibfnamefont{G.~M.} \bibnamefont{Gusev}},
  \bibinfo{author}{\bibfnamefont{O.~E.} \bibnamefont{Raichev}},
  \bibinfo{author}{\bibfnamefont{A.~K.} \bibnamefont{Bakarov}},
  \bibnamefont{and} \bibinfo{author}{\bibfnamefont{J.~C.}
  \bibnamefont{Portal}}, \bibinfo{journal}{Phys. Rev. B}
  \textbf{\bibinfo{volume}{84}}, \bibinfo{pages}{165303}
  (\bibinfo{year}{2011}).

\bibitem[{\citenamefont{Gusev et~al.}(2011)\citenamefont{Gusev, Wiedmann,
  Raichev, Bakarov, and Portal}}]{gusev:2011}
\bibinfo{author}{\bibfnamefont{G.~M.} \bibnamefont{Gusev}},
  \bibinfo{author}{\bibfnamefont{S.}~\bibnamefont{Wiedmann}},
  \bibinfo{author}{\bibfnamefont{O.~E.} \bibnamefont{Raichev}},
  \bibinfo{author}{\bibfnamefont{A.~K.} \bibnamefont{Bakarov}},
  \bibnamefont{and} \bibinfo{author}{\bibfnamefont{J.~C.}
  \bibnamefont{Portal}}, \bibinfo{journal}{Phys. Rev. B}
  \textbf{\bibinfo{volume}{83}}, \bibinfo{pages}{041306}
  (\bibinfo{year}{2011}).

\bibitem[{\citenamefont{Dmitriev et~al.}(2005)\citenamefont{Dmitriev, Vavilov,
  Aleiner, Mirlin, and Polyakov}}]{dmitriev:2005}
\bibinfo{author}{\bibfnamefont{I.~A.} \bibnamefont{Dmitriev}},
  \bibinfo{author}{\bibfnamefont{M.~G.} \bibnamefont{Vavilov}},
  \bibinfo{author}{\bibfnamefont{I.~L.} \bibnamefont{Aleiner}},
  \bibinfo{author}{\bibfnamefont{A.~D.} \bibnamefont{Mirlin}},
  \bibnamefont{and} \bibinfo{author}{\bibfnamefont{D.~G.}
  \bibnamefont{Polyakov}}, \bibinfo{journal}{Phys. Rev. B}
  \textbf{\bibinfo{volume}{71}}, \bibinfo{pages}{115316}
  (\bibinfo{year}{2005}).

\bibitem[{not({\natexlab{a}})}]{note:51}
\bibinfo{note}{In this notation, $\tq^{-1}=\tau_0^{-1}$ and
  $\ttr^{-1}=\tau_0^{-1}-\tau_1^{-1}$.}

\bibitem[{\citenamefont{Kozuka et~al.}(2014)\citenamefont{Kozuka, Tsukazaki,
  and Kawasaki}}]{kozuka:2014}
\bibinfo{author}{\bibfnamefont{Y.}~\bibnamefont{Kozuka}},
  \bibinfo{author}{\bibfnamefont{A.}~\bibnamefont{Tsukazaki}},
  \bibnamefont{and} \bibinfo{author}{\bibfnamefont{M.}~\bibnamefont{Kawasaki}},
  \bibinfo{journal}{Appl. Phys. Rev.} \textbf{\bibinfo{volume}{1}},
  \bibinfo{pages}{011303} (\bibinfo{year}{2014}).

\bibitem[{\citenamefont{Falson et~al.}(2015{\natexlab{a}})\citenamefont{Falson,
  Maryenko, Friess, Zhang, Kozuka, Tsukazaki, Smet, and
  Kawasaki}}]{falson:2015}
\bibinfo{author}{\bibfnamefont{J.}~\bibnamefont{Falson}},
  \bibinfo{author}{\bibfnamefont{D.}~\bibnamefont{Maryenko}},
  \bibinfo{author}{\bibfnamefont{B.}~\bibnamefont{Friess}},
  \bibinfo{author}{\bibfnamefont{D.}~\bibnamefont{Zhang}},
  \bibinfo{author}{\bibfnamefont{Y.}~\bibnamefont{Kozuka}},
  \bibinfo{author}{\bibfnamefont{A.}~\bibnamefont{Tsukazaki}},
  \bibinfo{author}{\bibfnamefont{J.~H.} \bibnamefont{Smet}}, \bibnamefont{and}
  \bibinfo{author}{\bibfnamefont{M.}~\bibnamefont{Kawasaki}},
  \bibinfo{journal}{Nat. Phys.} \textbf{\bibinfo{volume}{11}},
  \bibinfo{pages}{347} (\bibinfo{year}{2015}{\natexlab{a}}).

\bibitem[{\citenamefont{Falson et~al.}(2015{\natexlab{b}})\citenamefont{Falson,
  Kozuka, Smet, Arima, Tsukazaki, and Kawasaki}}]{falson:2015b}
\bibinfo{author}{\bibfnamefont{J.}~\bibnamefont{Falson}},
  \bibinfo{author}{\bibfnamefont{Y.}~\bibnamefont{Kozuka}},
  \bibinfo{author}{\bibfnamefont{J.~H.} \bibnamefont{Smet}},
  \bibinfo{author}{\bibfnamefont{T.}~\bibnamefont{Arima}},
  \bibinfo{author}{\bibfnamefont{A.}~\bibnamefont{Tsukazaki}},
  \bibnamefont{and} \bibinfo{author}{\bibfnamefont{M.}~\bibnamefont{Kawasaki}},
  \bibinfo{journal}{Appl. Phys. Lett.} \textbf{\bibinfo{volume}{107}},
  \bibinfo{eid}{082102} (\bibinfo{year}{2015}{\natexlab{b}}).

\bibitem[{\citenamefont{Kozlov et~al.}(2015)\citenamefont{Kozlov, Van'kov,
  Gubarev, Kukushkin, Solovyev, Falson, Maryenko, Kozuka, Tsukazaki, Kawasaki
  et~al.}}]{kozlov:2015}
\bibinfo{author}{\bibfnamefont{V.~E.} \bibnamefont{Kozlov}},
  \bibinfo{author}{\bibfnamefont{A.~B.} \bibnamefont{Van'kov}},
  \bibinfo{author}{\bibfnamefont{S.~I.} \bibnamefont{Gubarev}},
  \bibinfo{author}{\bibfnamefont{I.~V.} \bibnamefont{Kukushkin}},
  \bibinfo{author}{\bibfnamefont{V.~V.} \bibnamefont{Solovyev}},
  \bibinfo{author}{\bibfnamefont{J.}~\bibnamefont{Falson}},
  \bibinfo{author}{\bibfnamefont{D.}~\bibnamefont{Maryenko}},
  \bibinfo{author}{\bibfnamefont{Y.}~\bibnamefont{Kozuka}},
  \bibinfo{author}{\bibfnamefont{A.}~\bibnamefont{Tsukazaki}},
  \bibinfo{author}{\bibfnamefont{M.}~\bibnamefont{Kawasaki}},
  \bibnamefont{et~al.}, \bibinfo{journal}{Phys. Rev. B}
  \textbf{\bibinfo{volume}{91}}, \bibinfo{pages}{085304}
  (\bibinfo{year}{2015}).

\bibitem[{\citenamefont{Falson et~al.}(2016)\citenamefont{Falson, Kozuka,
  Uchida, Smet, Arima, Tsukazaki, and Kawasaki}}]{falson:2016}
\bibinfo{author}{\bibfnamefont{J.}~\bibnamefont{Falson}},
  \bibinfo{author}{\bibfnamefont{Y.}~\bibnamefont{Kozuka}},
  \bibinfo{author}{\bibfnamefont{M.}~\bibnamefont{Uchida}},
  \bibinfo{author}{\bibfnamefont{J.~H.} \bibnamefont{Smet}},
  \bibinfo{author}{\bibfnamefont{T.}~\bibnamefont{Arima}},
  \bibinfo{author}{\bibfnamefont{A.}~\bibnamefont{Tsukazaki}},
  \bibnamefont{and} \bibinfo{author}{\bibfnamefont{M.}~\bibnamefont{Kawasaki}},
  \bibinfo{journal}{Sci. Rep.} \textbf{\bibinfo{volume}{6}},
  \bibinfo{pages}{26598} (\bibinfo{year}{2016}).

\bibitem[{\citenamefont{K\"archer et~al.}(2016)\citenamefont{K\"archer,
  Shchepetilnikov, Nefyodov, Falson, Dmitriev, Kozuka, Maryenko, Tsukazaki,
  Dorozhkin, Kukushkin et~al.}}]{karcher:2016}
\bibinfo{author}{\bibfnamefont{D.~F.} \bibnamefont{K\"archer}},
  \bibinfo{author}{\bibfnamefont{A.~V.} \bibnamefont{Shchepetilnikov}},
  \bibinfo{author}{\bibfnamefont{Y.~A.} \bibnamefont{Nefyodov}},
  \bibinfo{author}{\bibfnamefont{J.}~\bibnamefont{Falson}},
  \bibinfo{author}{\bibfnamefont{I.~A.} \bibnamefont{Dmitriev}},
  \bibinfo{author}{\bibfnamefont{Y.}~\bibnamefont{Kozuka}},
  \bibinfo{author}{\bibfnamefont{D.}~\bibnamefont{Maryenko}},
  \bibinfo{author}{\bibfnamefont{A.}~\bibnamefont{Tsukazaki}},
  \bibinfo{author}{\bibfnamefont{S.~I.} \bibnamefont{Dorozhkin}},
  \bibinfo{author}{\bibfnamefont{I.~V.} \bibnamefont{Kukushkin}},
  \bibnamefont{et~al.}, \bibinfo{journal}{Phys. Rev. B}
  \textbf{\bibinfo{volume}{93}}, \bibinfo{pages}{041410}
  (\bibinfo{year}{2016}).

\bibitem[{\citenamefont{Falson et~al.}(2011)\citenamefont{Falson, Maryenko,
  Kozuka, Tsukazaki, and Kawasaki}}]{falson:2011}
\bibinfo{author}{\bibfnamefont{J.}~\bibnamefont{Falson}},
  \bibinfo{author}{\bibfnamefont{D.}~\bibnamefont{Maryenko}},
  \bibinfo{author}{\bibfnamefont{Y.}~\bibnamefont{Kozuka}},
  \bibinfo{author}{\bibfnamefont{A.}~\bibnamefont{Tsukazaki}},
  \bibnamefont{and} \bibinfo{author}{\bibfnamefont{M.}~\bibnamefont{Kawasaki}},
  \bibinfo{journal}{Appl. Phys. Express} \textbf{\bibinfo{volume}{4}},
  \bibinfo{pages}{091101} (\bibinfo{year}{2011}).

\bibitem[{\citenamefont{Hatke et~al.}(2009{\natexlab{b}})\citenamefont{Hatke,
  Zudov, Pfeiffer, and West}}]{hatke:2009a}
\bibinfo{author}{\bibfnamefont{A.~T.} \bibnamefont{Hatke}},
  \bibinfo{author}{\bibfnamefont{M.~A.} \bibnamefont{Zudov}},
  \bibinfo{author}{\bibfnamefont{L.~N.} \bibnamefont{Pfeiffer}},
  \bibnamefont{and} \bibinfo{author}{\bibfnamefont{K.~W.} \bibnamefont{West}},
  \bibinfo{journal}{Phys. Rev. Lett.} \textbf{\bibinfo{volume}{102}},
  \bibinfo{pages}{066804} (\bibinfo{year}{2009}{\natexlab{b}}).

\bibitem[{\citenamefont{Ando}(1974)}]{ando:1974b}
\bibinfo{author}{\bibfnamefont{T.}~\bibnamefont{Ando}}, \bibinfo{journal}{J.
  Phys. Soc. Jpn.} \textbf{\bibinfo{volume}{37}}, \bibinfo{pages}{1233}
  (\bibinfo{year}{1974}).

\bibitem[{\citenamefont{Laikhtman and Altshuler}(1994)}]{laikhtman:1994}
\bibinfo{author}{\bibfnamefont{B.}~\bibnamefont{Laikhtman}} \bibnamefont{and}
  \bibinfo{author}{\bibfnamefont{E.~L.} \bibnamefont{Altshuler}},
  \bibinfo{journal}{Ann. Phys. (N.Y.)} \textbf{\bibinfo{volume}{232}},
  \bibinfo{pages}{332} (\bibinfo{year}{1994}).

\bibitem[{\citenamefont{Giuliani and Quinn}(1982)}]{giuliani:1982}
\bibinfo{author}{\bibfnamefont{G.~F.} \bibnamefont{Giuliani}} \bibnamefont{and}
  \bibinfo{author}{\bibfnamefont{J.~J.} \bibnamefont{Quinn}},
  \bibinfo{journal}{Phys. Rev. B} \textbf{\bibinfo{volume}{26}},
  \bibinfo{pages}{4421} (\bibinfo{year}{1982}).

\bibitem[{not({\natexlab{b}})}]{note:8}
\bibinfo{note}{At $0.7 {\rm~T} < B < 1.5 {\rm~T}$, $ \pi \edc^2 \tin/\ttr \ll
  k_BT/\hbar\oc \ll \sqrt{\oc\ttr}$.}

\end{thebibliography}
\end{document}